\documentclass[aps,prl,twocolumn,groupedaddress,showpacs]{revtex4-1}
\pdfoutput=1

\usepackage{bm}
\usepackage{graphicx}
\usepackage{color}

\begin{document}

\title{Effects of Ru Substitution on Dimensionality and Electron Correlations in Ba(Fe$_{1-x}$Ru$_x$)$_2$As$_2$}

\author{
    N. Xu,$^1$
    T. Qian,$^1$
    P. Richard,$^1$
    Y.-B. Shi,$^1$
    X.-P. Wang,$^1$
    P. Zhang,$^1$
    Y.-B. Huang,$^1$
    Y.-M. Xu,$^2$
    H. Miao,$^1$
    G. Xu,$^1$
    G.-F. Xuan,$^3$
    W.-H. Jiao,$^3$
    Z.-A. Xu,$^3$
    G.-H. Cao,$^3$
    H. Ding,$^{1\ast}$
    }

\affiliation{$^1$Beijing National Laboratory for Condensed Matter Physics, and Institute of Physics, Chinese Academy of Sciences, Beijing 100190, China}
\affiliation{$^2$Materials Sciences Division, Lawrence Berkeley National Laboratory, Berkeley, California 94720, USA}
\affiliation{$^3$State Key Lab of Silicon Materials and Department of Physics, Zhejiang University, Hangzhou 310027, China}

\date{\today}

\begin{abstract}
We report a systematic angle-resolved photoemission spectroscopy study on Ba(Fe$_{1-x}$Ru$_x$)$_2$As$_2$ for a wide range of Ru concentrations (0.15 $\leq$ \emph{x} $\leq$ 0.74).  We observed a crossover from two-dimension to three-dimension for some of the hole-like Fermi surfaces with Ru substitution and a large reduction in the mass renormalization close to optimal doping. These results suggest that isovalent Ru substitution has remarkable effects on the low-energy electron excitations, which are important for the evolution of superconductivity and antiferromagnetism in this system.
\end{abstract}


\maketitle

Superconductivity in the iron-based materials usually emerges from a magnetic state by several kinds of routes leading to very similar phase diagrams of magnetism and superconductivity. In Ba$_{1-x}$K$_x$Fe$_2$As$_2$ \cite{BaK122} and Ba(Fe$_{1-x}$Co$_x$)$_2$As$_2$ \cite{Co122}, the introduction of extra hole or electron carriers shifts the chemical potential so that the sizes of the hole and electron Fermi surface (FS) pockets evolve oppositely \cite{Neupane_chemical_potential}, which eventually suppresses the nesting between the hole and electron FS pockets that play a role in the formation of spin-density-wave (SDW) with exotic Dirac cone dispersion \cite{Richard_cone} in the parent compound. While it is generally believed that external pressure also changes the FS topology by modifying the chemical bonds \cite{Ba122 pressure}, the role of isovalent element substitution is still debated. Various scenarios, for example changes of the FS topology by chemical pressure \cite{Canfield, XuZA Eu122,Colson}, the reduction of electron correlations \cite{SiQM,Colson}, magnetic dilution \cite{Kaminski}, and even the addition of extra hole carriers \cite{Feng P122}, have been suggested to explain the suppression of the SDW order with isovalent element substitution in the BaFe$_2$(As$_{1-x}$P$_x$)$_2$ and Ba(Fe$_{1-x}$Ru$_x$)$_2$As$_2$ systems. Surprisingly, only little attention has been devoted to answer the reversed but somehow similarly important question: how does superconductivity is suppressed by increasing the substitution further than the optimal concentration?

Since single-crystals can be grown for the entire phase diagram, the Ba(Fe$_{1-x}$Ru$_x$)$_2$As$_2$ system is ideal to investigate the suppression of the SDW order, the emergence of superconductivity and its disappearance with isovalent-substitution. We expect that the electronic structure near the Fermi level (\emph{E}$_F$) be substantially modified by the Ru substitution. Indeed, the isovalent Ru substitution at the Fe site leads to an anisotropic lattice distortion, resulting in a strong increase of the As-Fe(Ru)-As bond angle and a decrease of the As height from the Fe(Ru) plane \cite{Sharma, Rullier, Canfield}. The Hall coefficient, which is always negative and decreases with decreasing temperature in the parent compound BaFe$_2$As$_2$, increases with decreasing temperature in the Ru-substituted samples and even changes sign for large Ru concentrations \cite{Rullier}. With its capacity to resolve dispersive electronic states in the vicinity of $E_F$, angle-resolved photoemission spectroscopy (ARPES) is a powerful tool to determine which parameters drive the system from a SDW order in BaFe$_2$As$_2$ to a non-magnetic metallic state in BaRu$_2$As$_2$, with superconductivity occurring at mixed composition. However, not only previous ARPES studies did not characterized the suppression of superconductivity in the overdoped regime, the conclusions drawn on the suppression of the SDW order show a striking contrast. While Brouet \emph{et al.} observed a decrease in the electronic correlations and significant modifications of the Fermi surface with Ru substitution, Dhaka \emph{et al.} reported almost no changes in the FS topology and the band dispersions, which points towards a non-Fermi surface-driven mechanism such as magnetic dilution for the origin of the SDW suppression \cite{Kaminski}.



In this letter, we report ARPES results for a Ru substitution range wider than in previous studies. Our data indicate that the impact of the degradation of the nesting between hole and electron FS pockets cannot be neglected. However, the most significant changes in the electronic band structure are observed at optimal concentration and at higher Ru content. We reveal a clear crossover from 2D to 3D for some of the hole-like FSs with Ru substitution accompanied by a reduction of mass renormalization, suggesting that the isovalent Ru substitution has significant effects on the low-energy electron excitations. The implications of the changes in the FS topology and the near-\emph{E}$_F$ band dispersions on the evolution of the SDW order and superconductivity are discussed.

High quality single crystals of Ba(Fe$_{1-x}$Ru$_x$)$_2$As$_2$ were grown by the flux method \cite{sample}. ARPES experiments were performed at beamlines PGM and Apple-PGM of the Synchrotron Radiation Center (WI), beamline 12.0.1 of the Advanced Light Source (CA), and beamline SIS of the Swiss Light Source (Switzerland) with Scienta analyzers. The energy and angular resolutions were set at 20 meV and 0.2$^{\circ}$, respectively. The samples were cleaved \emph{in situ} and measured at 20 K in a vacuum better than 4$\times$10$^{-11}$ Torr.

\begin{figure}[!t]
\begin{center}
\includegraphics[width=3.4in]{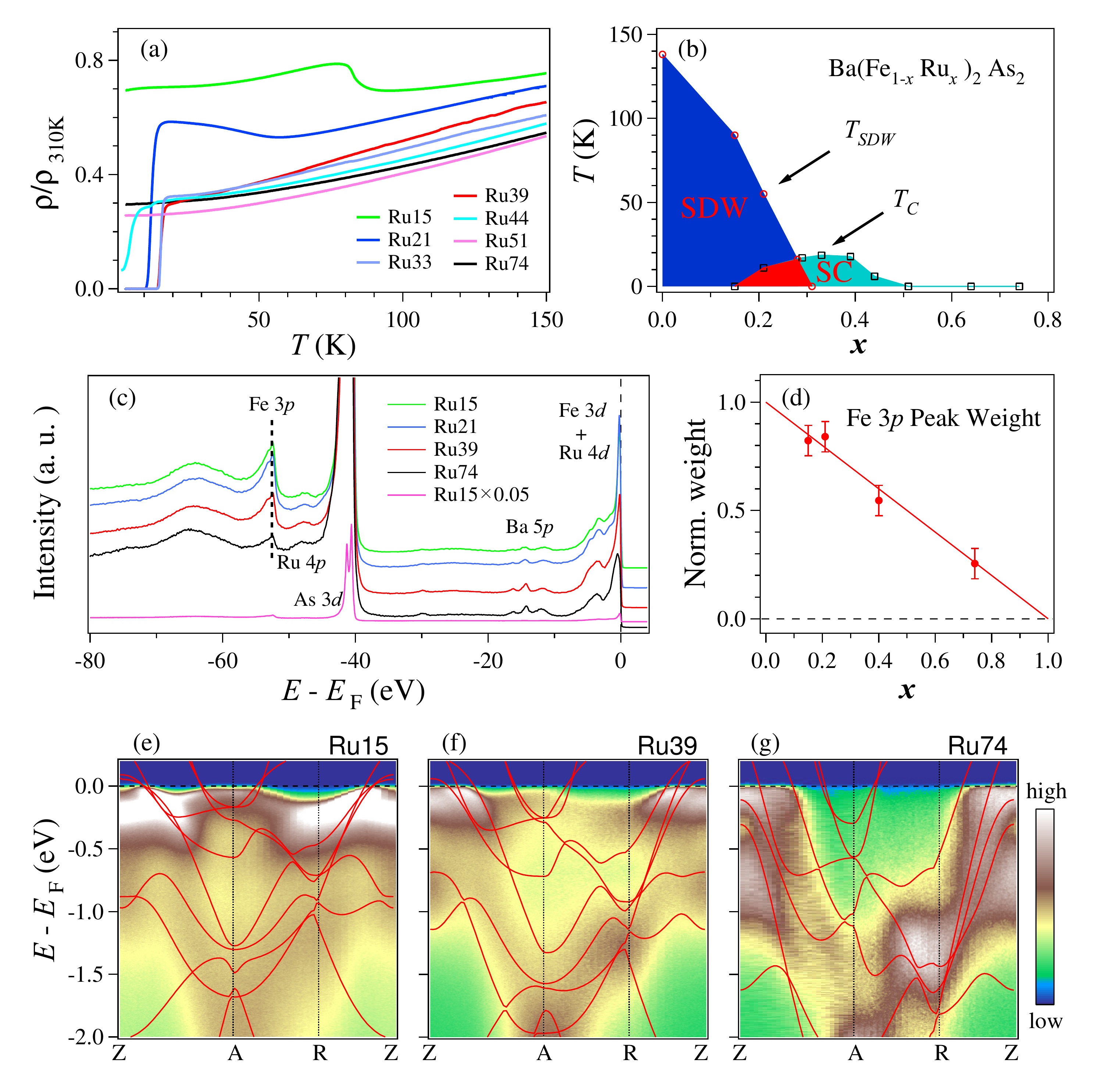}
\end{center}
\caption{(Color online): (a) Temperature dependence of the in-plane resistivity of Ba(Fe$_{1-x}$Ru$_x$)$_2$As$_2$ single crystals  for x=0.15, 0.21, 0.33, 0.39, 0.44, 0.51, and 0.74 respectively (named Ru15, Ru21, Ru33, Ru39, Ru44, Ru51 and Ru74). (b) Phase diagram of  Ba(Fe$_{1-x}$Ru$_x$)$_2$As$_2$ based on the resistivity data. (c) Core levels recorded with a photon energy of 195 eV. (d) Doping dependence of the Fe 3\emph{p} (E$_ b$ $\sim$ 53 eV) peak weight. (e)-(g) Intensity plot along Z-A-R-Z (\emph{k}$_z$=$\pi$) for Ru15, Ru39 and Ru74. Non-renormalized LDA bands are also plotted for comparison.}
\label{fig1}
\end{figure}

The normalized resistivity data plotted in Fig. 1(a) suggest that the SDW order identified as a resistive anomaly (upturn) is gradually shifted to low temperature with Ru substitution and disappears for \emph{x} $\geq$ 0.30. Superconductivity emerges for \emph{x} $>$ 0.15, reaches its maximum transition temperature at \emph{x} $\sim$ 0.30, and is strongly suppressed for \emph{x} $>$ 0.39. The SDW temperature \emph{T}$_{SDW}$ and critical temperature \emph{T}$_c$ extracted from the resistivity data are plotted in the \emph{T}-\emph{x} phase diagram, Fig. 1(b).  Fig. 1(c) shows the shallow core levels and the valence band spectra for different Ru concentrations. The peak position of the Fe 3\emph{p} orbital at the binding energy (\emph{E}$_B$) of $\sim$ 53 eV is not shifted with Ru substitution, confirming that the valence state of the Fe ions is not changed. Furthermore, the intensity of the Fe 3\emph{p} peak extracted by integrating the peak area decreases linearly as a function of the Ru concentration and can be extrapolated to zero for BaRu$_2$As$_2$, as shown in Fig. 1(d). In contrast, the Ru 4\emph{p} peak at \emph{E}$_B$ $\sim$ 48 eV is enhanced with increasing the Ru concentration, though it is too weak to extract accurately its intensity value. 

In iron-based superconductors, most of dispersive bands originating from the Fe 3\emph{d} orbitals are located within 2 eV below \emph{E}$_F$. To illustrate the effects of Ru substitution on the dispersive bands, we plot in Figs. 1(e)-1(g) the photoemission intensity along the high symmetry lines Z-A-R-Z for \emph{x} = 0.15, 0.39 and 0.74, respectively. For comparison, we also plot the band structure within the local density approximation (LDA) for \emph{x} = 0, 0.4, and 1 in Figs. 1(e)-1(g), respectively. While the overall bandwidth increases slightly for higher Ru concentrations as expected from band calculations, it is interesting to notice that the calculated bands do not need to be renormalized to fit the dispersive bands in the high \emph{E}$_B$ range. In contrast, the near-\emph{E}$_F$ bands are strongly renormalized, especially for $x\leq 0.39$, suggesting that the many interactions mainly affect the low-energy electron excitations.

\begin{figure}[!t]
\begin{center}
\includegraphics[width=3.4in]{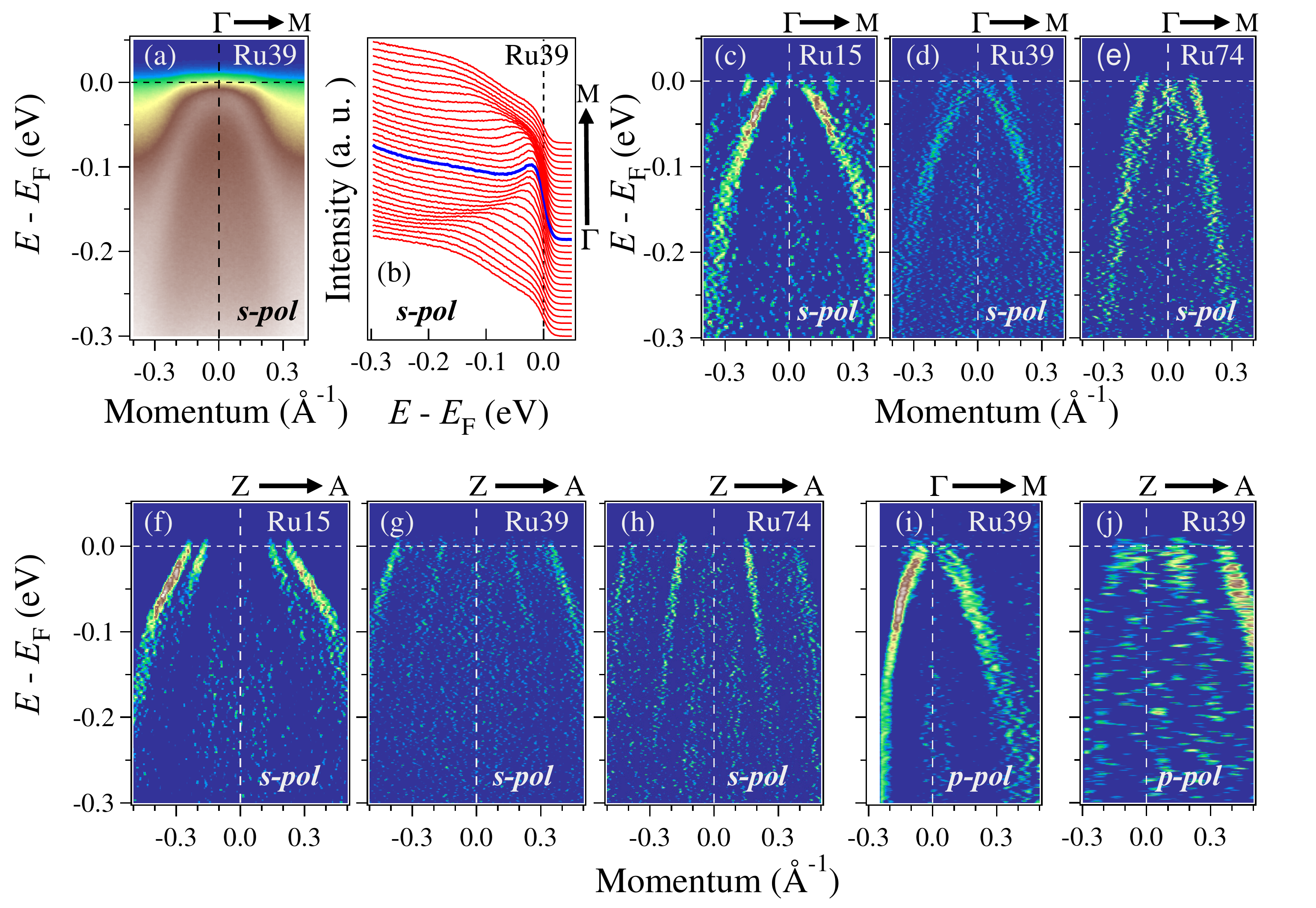}
\end{center}
\caption{(Color online) (a) ARPES intensity plot along $\Gamma$-M (\emph{k}$_z$=0) for Ru15, taken in the \emph{s}-\emph{pol} geometry. (b) Corresponding energy distribution curves. (c)-(e) Curvature plot of the momentum distribution curves (MDCs) along $\Gamma$-M (\emph{k}$_z$=0) in the \emph{s}-\emph{pol} geometry. (f)-(h)  Same as (c)-(e), but along the Z-A (\emph{k}$_z$=$\pi$) symmetry line. (i) Same as (d), but in \emph{p}-\emph{pol} geometry. (i) Same as (g), but in \emph{p}-\emph{pol} geometry.}
\label{fig2}
\end{figure}

In order to clarify the effects of Ru substitution on the low-energy excitations, we focus on the band dispersions near \emph{E}$_F$. Fig. 2(a) shows a photoemission intensity plot at the $\Gamma$ point (\emph{k}$_z$ = 0 at the BZ center) in \emph{s} polarization geometry (\emph{s}-\emph{pol}) for \emph{x} = 0.39. Fig. 2(b) and Fig. 2(d) show the corresponding EDCs and the corresponding curvature plot of the momentum distribution curves (MDCs) \cite{curvature}, respectively. One can see that the curvature plot exhibits more clearly the band dispersions near \emph{E}$_F$. We show the curvature plots of the MDCs at the $\Gamma$ and Z (\emph{k}$_z$ = $\pi$) points for \emph{x} = 0.15, 0.39, and 0.74 in \emph{s}-\emph{pol} geometry in Figs. 2(c)-2(h), respectively. As seen in Fig. 2(c), two hole-like bands crossing \emph{E}$_F$ are observed at $\Gamma$ for \emph{x} = 0.15. As the Ru content increases, their Fermi momenta  (\emph{k}$_F$'s) become smaller and the inner band even sinks below \emph{E}$_F$ for \emph{x} = 0.39 and 0.74. The effect of Ru substitution is somehow opposite at the Z point. Not only the two hole-like bands cross \emph{E}$_F$ at Z for all samples, their respective \emph{k}$_F$'s actually increase with the Ru concentration. These results indicate that Ru substitution changes substantially the near-\emph{E}$_F$ band structure. In addition, we observed the third predicted hole-like band at the BZ center in the \emph{p} polarization geometry (\emph{p}-\emph{pol}). As seen in Figs. 2(i) and 2(j), the band sinks below \emph{E}$_F$ at $\Gamma$ but crosses \emph{E}$_F$ at Z for \emph{x} = 0.39.

\begin{figure}[!t]
\begin{center}
\includegraphics[width=3.4in]{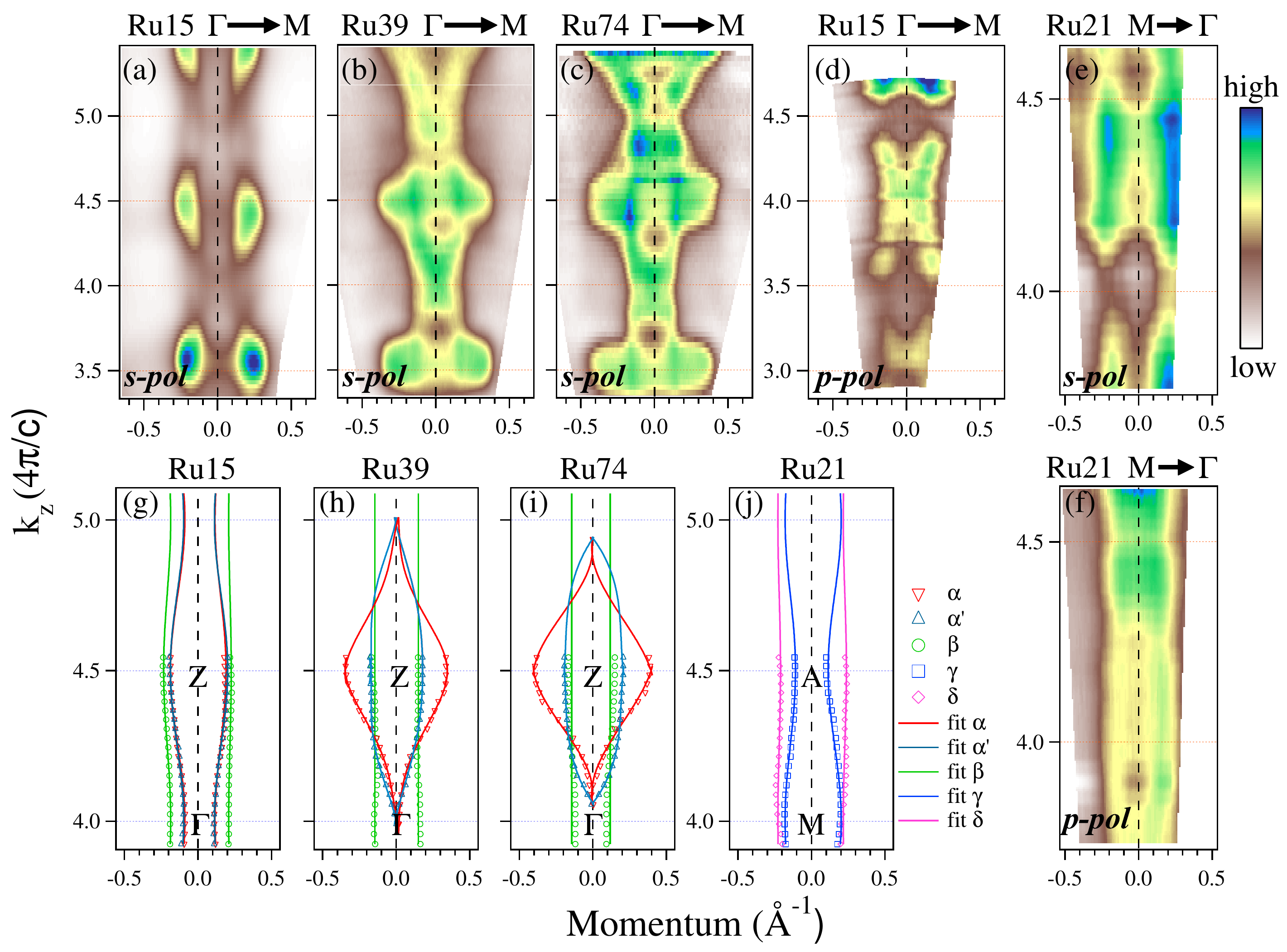}
\end{center}
\caption{(color online)(a)-(c) Hole FS maps in \emph{k}$_x$-\emph{k}$_z$ plane in \emph{s}-\emph{pol} geometry. (d) Same as (a), but in the \emph{p}-\emph{pol} geometry. (e) Electron FS maps in  \emph{k}$_x$-\emph{k}$_z$ plane for Ru21 in the \emph{s}-\emph{pol} geometry. (f) Same as (e), but obtained in the \emph{p}-\emph{pol} geometry. (g)-(i) Fermi momenta (symbols) of the hole bands. (j) Fermi momenta of the electron bands (symbols).}
\label{fig3}
\end{figure}

To further clarify the evolution of the FS topology in the 3D BZ with Ru substitution, we performed ARPES measurements at different \emph{k}$_z$ by varying the photon energy. Figs. 3(a)-3(c) show the intensity plots integrated within $\pm$10 meV with respect to \emph{E}$_F$ in the \emph{k}$_x$-\emph{k}$_z$ plane using the \emph{s}-\emph{pol} geometry, for \emph{x} = 0.15, 0.39 and 0.74, respectively. One can see two FSs clearly in the \emph{s}-\emph{pol} geometry. One of them (hereafter called $\beta$) keeps a quasi-2D behavior for all Ru concentrations whereas the other one (called $\alpha$) exhibits clearly a crossover from 2D to 3D with Ru substitution. Fig. 3(d) shows the \emph{p}-\emph{pol} geometry data for Ru15, in which one can see a third FS (called $\alpha$'). As shown in Figs. 3(e) and 3(f), two quasi-2D electron-like FSs are distinguished at the M point in both the \emph{s}-\emph{pol} and \emph{p}-\emph{pol} geometries, with no obvious change upon Ru substitution. 

\begin{figure}[!t]
\begin{center}
\includegraphics[width=3.4in]{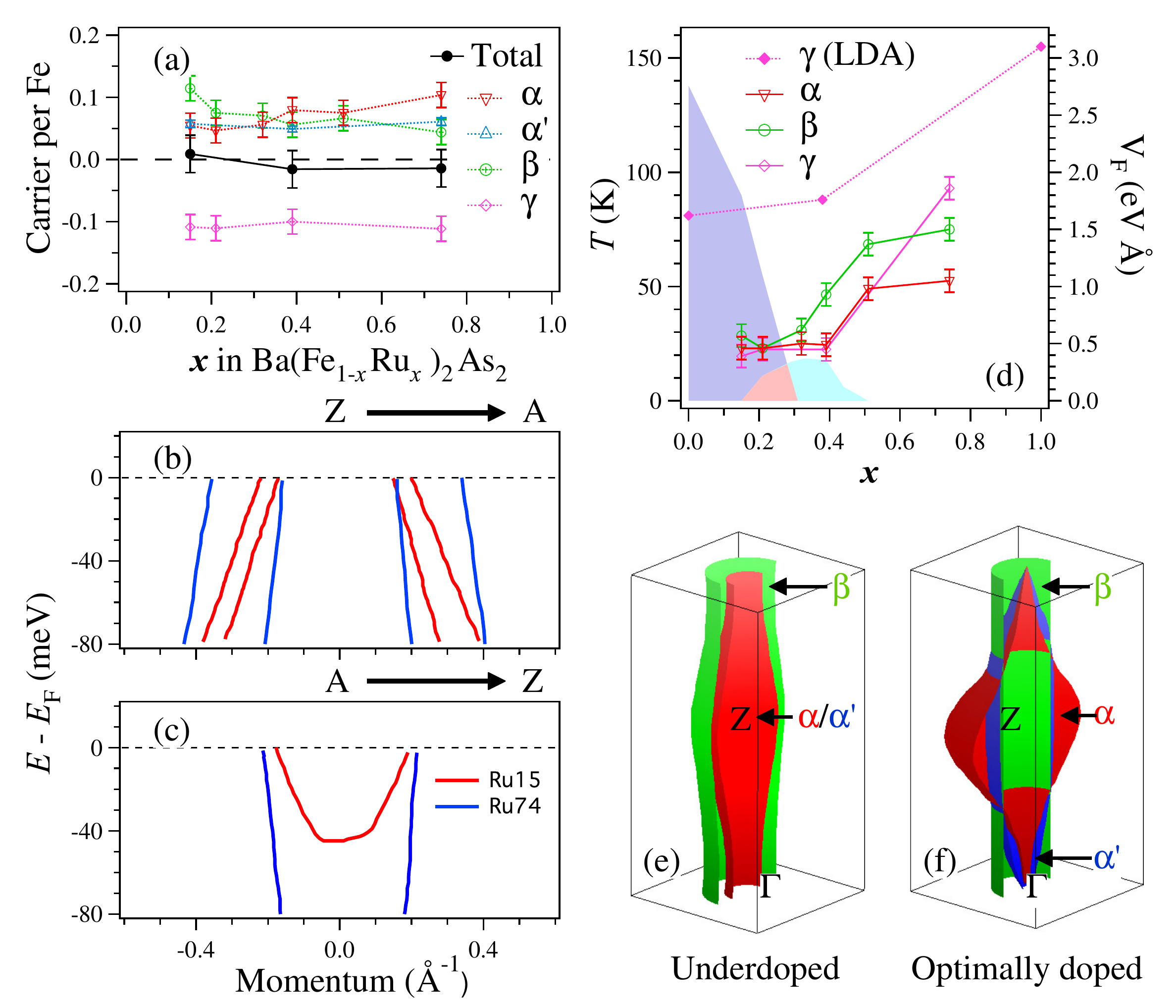}
\end{center}
\caption{(a) Total carrier concentration (black full symbols) deduced from the contribution of individual bands (open symbols). (b)-(c) Band dispersions obtained from MDC fits of the hole and electron pockets \emph{k}$_z$=$\pi$), respectively, for Ru15 and Ru74. (d) Doping dependence of the Fermi velocity of  the $\alpha$ band (red line), $\beta$ band (green line) and $\gamma$ band (pink line) at the Z point. The full pink diamonds correspond to the Fermi velocity of the  $\gamma$ band predicted by our LDA calculations. The phase diagram is also plotted for convenience. (e)-(f) 3D hole FS plots for underdoped and optimally-doped Ba(Fe$_{1-x}$Ru$_x$)$_2$As$_2$.}
\label{fig4}
\end{figure}

The \emph{k}$_F$'s of the three hole bands extracted from the MDCs are plotted in Figs. 3(g)-3(i) while the \emph{k}$_F$'s for the two electron bands are given in Fig. 3(j). From these extracted 3D \emph{k}$_F$ values, we can estimate the number of carriers associated with each FS pocket. The results, as well as the total number of carriers, are displayed as a function of Ru concentration in Fig. 4(a). The hole carriers are well compensated by the electron ones at all Ru concentrations, demonstrating that the Ru substitution is isovalent, in agreement with the core-level measurements and previous ARPES studies \cite{Kaminski,Colson}. Although the carrier concentration does not change with Ru substitution, Fig. 4(a) clearly shows that the band structure evolves. In particular, there is a hole carrier transfer from the $\beta$ band to the $\alpha$ band that occurs from low doping to higher doping, suggesting that the nesting conditions between the $\Gamma$-centered hole FS pockets and M-centered electron FS pockets are affected as well. Although other factors such as magnetic dilution should be taken in account \cite{Kaminski}, our experimental results thus suggest that the degradation of the nesting conditions may play a role in the suppression of the SDW ordering with Ru increasing.  

To further illustrate the effects of the Ru substitution on the low-energy excitations, we compare the extracted dispersions of the hole- and the electron-like bands at \emph{k}$_z$ = $\pi$ for \emph{x} = 0.15 and 0.74 in Figs. 4(b) and 4(c), respectively. The dispersions of all the bands for \emph{x} = 0.74 are much steeper than those for \emph{x} = 0.15. Fig. 4(d) displays the extracted Fermi velocity (\emph{v}$_F$) values at \emph{k}$_z$ = $\pi$ as a function of Ru concentration. While the \emph{v}$_F$ values in the $k_z=\pi$ plane are not obviously changed in the underdoped range, they increase rapidly in the overdoped range. This effect is captured by our LDA band structure calculations, for which the $v_F$ values corresponding to the $\gamma$ band in the $k_z=\pi$ plane are shown in Fig. 4(d). We note that these sudden changes in $v_F$ and in the band renormalization occur near optimal doping. From the phase diagram given in Fig. 4(d), one can see that the optimal doping for superconductivity is also close to the phase boundary of the SDW state. It is interesting to point out that similar trend in mass enhancement towards the boundary of the magnetic order phase have also been revealed in another isovalent substituted system, BaFe$_2$(As$_{1-x}$P$_x$)$_2$ \cite{dHvA3}, and in the heavy fermion system CeRhIn$_5$ by de Haas-van Alphen experiments \cite{HF}. This gives us a hint that the many body interactions, which give rise to the enhancement of the effective mass of the quasiparticles in the low doping regime, could be associated with the magnetic quantum critical point (QCP) and could play an important role in the unconventional superconductivity of these systems \cite{SiQM2}.


Whether a localized or an itinerant picture is most appropriate to describe superconductivity in the Fe-based superconductors is still a matter of controversy \cite{Richard_review}. Let's first assume that an itinerant picture prevails. In Figs. 4(e)-4(f), we plot the 3D hole FSs for underdoped and optimally doped Ba(Fe$_{1-x}$Ru$_x$)$_2$As$_2$. Our results show that the hole-like $\beta$ FS remains quasi-2D and is still quasi-nested with the electron-like FSs far beyond the superconducting dome. In contrast, the $\alpha$ and $\alpha$' FSs become very 3D with Ru substitution. The tops of the $\alpha$ and $\alpha$' bands are degenerated at $\Gamma$ [see Figs. 2(c) and 2(i)], suggesting that they mainly originate from the \emph{d}$_{xz/yz}$ orbitals, in agreement with a recent ARPES study on the (Ba,K)Fe$_2$As$_2$ system \cite{Xiaoping_orbital}. The enhancement of the 3D character is attributed to a stronger \emph{d}-\emph{p} hybridization of the transition metal atoms with As \cite{Singh} due to a larger spatial extension of the Ru 4\emph{d} orbitals and a decrease of the As height \cite{Sharma, Rullier, Canfield}. On the other hand, the quasi-2D $\beta$ FS is attributed to the \emph{d}$_{xy}$ orbital, which spatial extension is limited to the \emph{ab} plane. Our results indicate clearly that if quasi-nesting between the hole and electron FS pockets is the determinant factor for superconductivity in these materials, then the \emph{d}$_{xz/yz}$ orbitals are mainly responsible. Within a local picture, the main interpretation for the suppression of superconductivity at high Ru content can be explained in terms of both the reduction of the electronic correlations reported in this Letter and the important modifications of the local antiferromagnetic exchange constants due to the enhanced spacial extension of the Ru 4$d$ orbitals and their increased hybridization with the As 3$p$ orbitals due to the decrease of the As height. 

In summary, our study of Ba(Fe$_{1-x}$Ru$_x$)$_2$As$_2$ over a wide range of doping indicates that the degradation of the nesting conditions between the hole and electron FS pockets cannot be neglected in understanding the suppression of the SDW in this system. However, the main changes in the electronic band structure occur for Ru concentrations exceeding optimal substitution. In that regime, some hole bands become much more 3D, an effect accompanied by an increase of the Fermi velocities and a strong suppression of the electronic correlations. Our study suggests that one or many of these strong effects may play a role in the suppression of superconductivity in the high-substitution regime of Ba(Fe$_{1-x}$Ru$_x$)$_2$As$_2$.

We acknowledge X. Dai, Z. Fang, Z. Pan, W. Yin and P. C. Canfield for useful discussions. This work was supported by grants from CAS (2010Y1JB6), MOST (2010CB923000 and 2011CBA001000) and NSFC (11004232 and 11050110422). This work is based in part upon research conducted at the Synchrotron Radiation Center which is primarily funded by the University of Wisconsin-Madison with supplemental support from facility Users and the University of Wisconsin-Milwaukee.

\end{document}